\documentclass[aps,prl,amsmath,amssymb,superscriptaddress,twocolumn,10pt]{revtex4-2}

\usepackage[english]{babel}
\usepackage[utf8]{inputenc}
\usepackage{amsfonts}
\usepackage[T1]{fontenc}
\usepackage[pdftex]{graphicx}
\usepackage{hyperref}
\hypersetup{
    unicode=false,          % non-Latin characters in AcrobatÕs bookmarks
    pdftoolbar=false,        % show Acrobat's toolbar?
    pdfmenubar=true,        % show Acrobat's menu?
    pdffitwindow=false,     % window fit to page when opened
    pdfstartview={FitH},
    pdfnewwindow=true,      % links in new window
    colorlinks=true,       % false: boxed links; true: colored links
    linkcolor=black,          % color of internal links (change box color with linkbordercolor)
    citecolor=blue,        % color of links to bibliography
    filecolor=magenta,      % color of file links
    urlcolor=blue           % color of external links
}

\usepackage{url}
\usepackage[titletoc,title]{appendix}
\usepackage{epstopdf}
\usepackage{amsmath}
\usepackage{amssymb}
\usepackage{dsfont}
\usepackage{mathtools}
\usepackage{bbold}
\usepackage{xcolor}
\usepackage{tikz}
\usetikzlibrary{arrows.meta}
\usetikzlibrary{shapes.callouts,calc}

\graphicspath{{./Figures/}}

\newcommand{\tr}{{\rm tr}}

\newcommand{\ket}[1]{|#1\rangle}
\newcommand{\bra}[1]{\langle#1|}

\newcommand{\Sx}[1]{\hat S^x_{#1}}
\newcommand{\Sy}[1]{\hat S^y_{#1}}
\newcommand{\Sz}[1]{\hat S^z_{#1}}
\newcommand{\Sa}[1]{\hat S^\alpha_{#1}}
\newcommand{\SP}[1]{\hat S^+_{#1}}
\newcommand{\SM}[1]{\hat S^-_{#1}}
\newcommand{\Pup}[1]{\hat{\mathbb{P}}^\uparrow_{#1}}
\newcommand{\Pdn}[1]{\hat{\mathbb{P}}^\downarrow_{#1}}

\newcommand{\mytitle}{Collective Dynamics in Spin Chains with Constrained Dissipation: \\Classically Fragile but Quantum Robust}

\makeatletter
\newsavebox{\@brx}
\newcommand{\llangle}[1][]{\savebox{\@brx}{\(\m@th{#1\langle}\)}%
  \mathopen{\copy\@brx\kern-0.5\wd\@brx\usebox{\@brx}}}
\newcommand{\rrangle}[1][]{\savebox{\@brx}{\(\m@th{#1\rangle}\)}%
  \mathclose{\copy\@brx\kern-0.5\wd\@brx\usebox{\@brx}}}
\makeatother

\def\be{\begin{equation}}
\def\ee{\end{equation}}
\def\ba{\begin{eqnarray}}
\def\ea{\end{eqnarray}}

\newcommand{\ssz}
\begin{document}

\title{\mytitle}
\author{Pietro Brighi} %\orcidlink{0000-0002-2399-5827}}
\email{pietro.brighi@univie.ac.at}
\affiliation{
    Faculty of Physics, University of Vienna, Boltzmanngasse 5, 1090 Vienna, Austria
    }

\author{Gianluca Frazzei}
\affiliation{
    Pitaevskii BEC Center, CNR-INO and Dipartimento di Fisica, Universit\`a di Trento, I-38123 Trento, Italy
    }
\affiliation{
    INFN-TIFPA, Trento Institute for Fundamental Physics and Applications, I-38123 Trento, Italy
    }

\author{Igor Lesanovsky}
\affiliation{
    Institut für Theoretische Physik and Center for Integrated Quantum Science and Technology (IQST), Universität Tübingen, Auf der Morgenstelle 14, 72076 Tübingen, Germany
    }
\affiliation{
    School of Physics and Astronomy, University of Nottingham, University Park, Nottingham NG7 2RD, United Kingdom
    }
\affiliation{
    Centre for the Mathematics and Theoretical Physics of Quantum Non-Equilibrium Systems, University of Nottingham, Nottingham NG7 2RD, United Kingdom
    }

\author{Juan P. Garrahan}
\affiliation{
    School of Physics and Astronomy, University of Nottingham, University Park, Nottingham NG7 2RD, United Kingdom
    }
\affiliation{
    Centre for the Mathematics and Theoretical Physics of Quantum Non-Equilibrium Systems, University of Nottingham, Nottingham NG7 2RD, United Kingdom
    }

\author{Alberto Biella}
\affiliation{
    Pitaevskii BEC Center, CNR-INO and Dipartimento di Fisica, Universit\`a di Trento, I-38123 Trento, Italy
    }
\affiliation{
    INFN-TIFPA, Trento Institute for Fundamental Physics and Applications, I-38123 Trento, Italy
    }

\date{\today}   

\begin{abstract}
    Collective non-equilibrium phenomena in many-body systems are strongly influenced by fluctuations, particularly in the vicinity of phase transitions, where the interplay between classical and quantum effects can alter cooperative behavior. Here, we investigate this classical-quantum competition in a kinetically constrained spin chain, the dissipative far-East model. 
    In absence of fluctuations, the system exhibits emergent collective dynamics, resulting in phase coexistence. 
    We find that this phenomenon is robust to quantum fluctuations, which further stabilize it, whereas classical fluctuations suppress phase coexistence at long times.
    We map out the steady-state phase diagram using a cluster mean-field approach and characterize the dynamics with stochastic tensor-network methods. Within the bistable phase we identify signatures of dynamical heterogeneity in both the local magnetization and the entanglement dynamics. This highlights how competing quantum fluctuations and classical kinetic constraints shape the relaxation dynamics and the phase structure of many-body systems.
\end{abstract}

\maketitle

\textbf{\textit{Introduction--}} Open quantum many-body systems represent a premier frontier for exploring novel non-equilibrium phenomena (for recent reviews see Refs.~\cite{fazio2025many-body,sieberer2025universality,minganti-fabrizio2026open}). In these systems, coherent quantum dynamics and dissipation compete resulting in many-body phases, critical behaviors and collective phenomena without an equilibrium counterpart, see e.g.\ Refs.~\cite{diel2008quantum,verstraete2009quantum,sieberer2016keldysh,minganti2018spectral,jin2018phase,Biella2018,Biella2017,Iemini2018,Reyhaneh2025,Rosso2022}.

\begin{figure}[t]
    \centering
\includegraphics[width=0.9\linewidth]{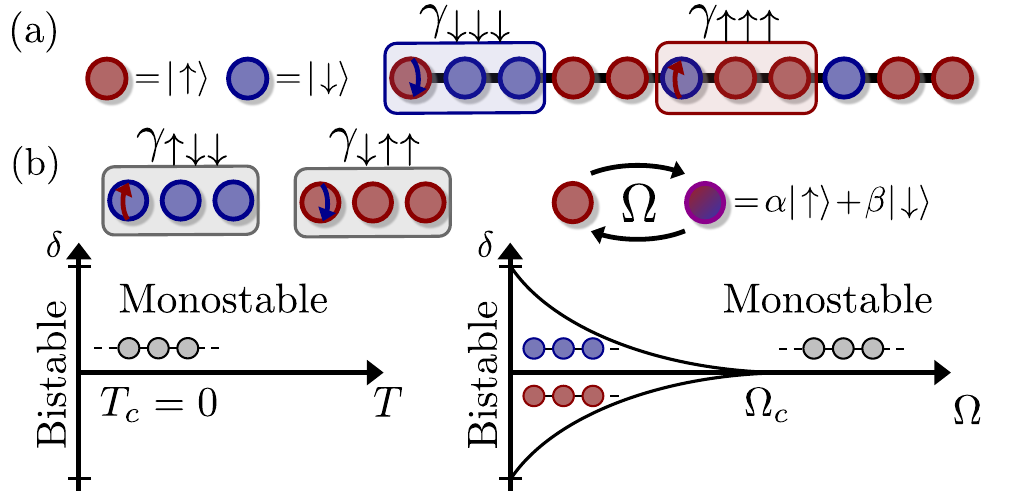}
    \caption{\label{Fig:sketch + bubble}
    (a) The far-East model:  dissipation acts on the left site of the three-sites plaquette, aligning it with the majority within the plaquette itself.
    (b) Sketch of the steady-phase diagram: classical fluctuations (of strength $T=(\gamma_{\downarrow\uparrow\uparrow}+\gamma_{\uparrow\downarrow\downarrow})/\gamma$) allow to flip the spin against the majority of the plaquette, violating the kinetic constraint. 
    The transverse field $\Omega$ induces a coherent superposition of the up and down state.
    Bistability is fragile with respect to classical fluctuations, yielding a monostable phase for any $T>0$.
    On the contrary, quantum fluctuations stabilize collective behaviors leading to an extended bistable region (up to a finite critical field $\Omega_c/\gamma\approx0.15$ at $\delta=0$). 
    }
\end{figure}

The dissipative, incoherent, part of the dynamics can be designed to encode particular classical non-equilibrium processes~\cite{gambetta2019discrete,carollo2022signatures}. One such class corresponds to kinetically constrained models (KCMs) \cite{fredrickson1984kinetic,ritort2003glassy,garrahan2011kinetically,garrahan2018aspects,hartarsky2025kinetically}, that is, classical stochastic systems with explicit local constraints in their dynamics. A concrete example is the East model~\cite{jackle1991a-hierarchically}, a one-dimensional lattice spin system where the ability of a spin to flip is determined by the state of one of its two neighbours (the one to the ``East''). The classical East model and its variants have proven important in the study of the glass transition \cite{garrahan2002geometrical,garrahan2007dynamical,chandler2010dynamics,hasyim2023emergent} and more broadly, through its quantum generalisations, in the study of quantum non-ergodicity \cite{horssen2015dynamics,pancotti2020quantum,roy2020strong,brighi2023quantumfragm,bertini2024exact,bertini2024localized,brighi2024anomalous,menzler2025graph,lazarides2026locality,Paviglianiti2025}. The implementation of other kinds of models with constrained dynamical rules, such as cellular automata, 
on quantum hardware~\cite{huerta-alderete2020quantum,jones2022small-world,suprano2024photonic,white2026quantum}, has inspired further quantum extensions~\cite{farrelly2020a-review,p.2019an-overview}, with applications to quantum dynamics~\cite{gambetta2019discrete,gillman2021quantum} and algorithmics~\cite{raussendorf2005quantum,guedes2024quantum,wagner2025density}.

Within a master equation formalism one can define KCMs in which classical and quantum processes compete. Quantum unitary evolution can enrich the behavior of purely dissipative systems, and the competiton of coherent and incoherent processes can give rise to novel forms of collective behavior and non-equilibrium phases~\cite{carollo2022signatures,perfetto2023reaction-limited,lan2018quantum,causer2025dynamical}. For the case of non-equilibrium phase transitions, what is less well understood is what role competing fluctuations play in stabilising ordered phases. The expected behavior~\cite{toom1974nonergodic,bennett1985role,kazemi2021genuine,brighi2026dissipative} is that in two-dimensional lattice systems thermal and quantum effects can cooperate to stabilise transitions from disordered to ordered states, 
while in one dimension fluctuations are too strong, making ordered states fragile to them. Finding examples where this is not the case, and specifically where quantum dynamics can stabilize behaviors which would be destroyed by classical fluctuations, is therefore a timely question. 

In this Letter we address this problem by considering a quantum dissipative KCM consisting of a spin chain where dissipation obeys a chiral update rule involving the (right) nearest- and next-to-nearest- neighbors, supplemented with local Hamiltonian dynamics and/or classical noise. In the classification of KCMs this would be a two-spin facilitated model with an asymmetric constraint \cite{ritort2003glassy,hartarsky2025kinetically,sfairopoulos2026multicriticality}, and we call it the far-East model (fEM) to distinguish it from the standard East model (which is asymmetric one-spin facilitated). Investigating its behavior, we show that in the fEM quantum fluctuations can stabilize collective behavior which would only be transient in the presence of classical fluctuations alone. This results in a genuine non-equilibrium stationary state phase transition from a disordered (normal or paramagnetic) phase, to an ordered (bistable or ferromagnetic) phase made stable by quantum fluctuations.

To probe this physics, we employ cluster expansions~\cite{jin2016cluster,salatino2026} to determine the steady-state phase diagram, alongside tensor-network techniques~\cite{vidal2003efficient,schollwock2011the-density-matrix,orus2014a-practical,maki2023monte} to investigate the dynamics for system sizes beyond any exact numerical methods. 
We show that while classical fluctuations lead to the breakdown of bistability, when only quantum fluctuations are present a finite bistable region in parameter space survives. 
As is typical of classical KCMs \cite{garrahan2002geometrical}, relaxation is spatially heterogeneous, here as a consequence of the interplay between phase coexistence and kinetic constraints. A genuine quantum feature is the dynamical heterogeneity in bipartite entanglement, which is also seen in other (but non-dissipative) quantum KCMs~\cite{lan2018quantum,zadnik2023slow}.

\textbf{\textit{Model--}} 
The fEM is a driven-dissipative one-dimensional spin-$1/2$ chain of $N$ sites governed by a Lindblad master equation (hereafter we set $\hbar=1$)
\be
\label{eq:LindME}
\dot \rho = -i \left[\hat{H},\rho\right] + \sum_{j,\alpha} \left(\hat{L}^{\alpha}_j \rho (\hat{L}^{\alpha}_j)^\dagger - \frac12 \left\{(\hat{L}^{\alpha}_j)^\dagger \hat{L}^{\alpha}_j, \rho\right\}\right).
\ee
Dissipation is encoded into the set of local jump operators $\{\hat{L}^{\alpha}_j \}$ and induces a \textit{constrained action}, as sketched in Fig.~\ref{Fig:sketch + bubble}(a).
Specifically, the spin at site $j$ flips to align with the spins at sites $j+1$ and $j+2$, provided they are mutually aligned.
This is achieved introducing the following jump operators
\begin{align}
    \label{Eq:Lnu}
    \hat{L}^{\uparrow\uparrow\uparrow}_j &= \sqrt{\gamma_{\uparrow\uparrow\uparrow}}\SP{j}\Pup{j+1}\Pup{j+2}, \\
    \label{Eq:Lmu}
    \hat{L}^{\downarrow\downarrow\downarrow}_j &= \sqrt{\gamma_{\downarrow\downarrow\downarrow}}\SM{j}\Pdn{j+1}\Pdn{j+2},
\end{align}
where $\hat{S}^{\pm}_{j} = \Sx{j} \pm \imath\Sy{j}$ are the spin raising and lowering operators, $\hat{\mathbb{P}}^{\uparrow \downarrow}_{j} = \frac{1}{2}\pm\Sz{j}$ are the projectors onto the up and down state, respectively, and $\{\Sa{j},\alpha=x,y,z \}$ are the local spin matrices (with eigenvalues $\pm1/2$). We parametrize the rates as $\gamma_{\uparrow\uparrow\uparrow} = \gamma {(1+\delta)/}{2},\quad \gamma_{\downarrow\downarrow\downarrow} = \gamma {(1-\delta)/}{2}$, with $\gamma = \gamma_{\uparrow\uparrow\uparrow}+\gamma_{\downarrow\downarrow\downarrow}\in\mathbb{R}^+$ defining the overall dissipation strength, and $\delta \in[-1,1]$ playing the role of a bias favouring positive or negative magnetization.

We introduce classical fluctuations implemented by the jump operators $\hat{L}^{\downarrow\uparrow\uparrow}_j = \sqrt{\gamma_{\downarrow\uparrow\uparrow}}\SM{j}\Pup{j+1}\Pup{j+2}$ and $\hat{L}^{\uparrow\downarrow\downarrow}_j = \sqrt{\gamma_{\uparrow\downarrow\downarrow}}\SP{j}\Pdn{j+1}\Pdn{j+2}$, which control the ``softness'' of the constraint.
The strength of these processes is quantified through $T = {(\gamma_{\downarrow\uparrow\uparrow}+\gamma_{\uparrow\downarrow\downarrow})/}{\gamma}$. When $T=0$ the fEM is a KCM with ``hard'' constraints, while for small $T>0$ it is one with ``soft'' constraints~\cite{ritort2003glassy,elmatad2010finite-temperature}, where the kinetic constraint can be violated occasionally. In the End Matter we study the case $T>0$, but in the rest of the main text we only consider the fEM at $T=0$ where these classical constraint-breaking processes are absent.
Quantum coherent processes are introduced through the Hamiltonian $\hat{H} = \Omega\sum_j \Sx{j}$, which creates local superposition of $\uparrow$ and $\downarrow$ configurations with an amplitude given by the transverse field $\Omega$.

\textbf{\textit{Steady-state phase diagram--}} Figure~\ref{Fig:sketch + bubble}(b) shows the main features of the steady-state phase diagram. 
When the constraint is hard and in absence of quantum fluctuations, the dissipators~\eqref{Eq:Lnu} and \eqref{Eq:Lmu} fully determine the dynamics and allow for the emergence of multiple stationary states [such that $\dot \rho=0$ in Eq.~\eqref{eq:LindME}] whose number depends on $N$ and on the boundary conditions \footnote{In open boundary conditions, the dissipators support $2N$ stationary states. Under periodic boundary conditions, for odd $N$ only the $2$ polarized states are stationary, while for even $N$ the $2$ N\'eel states are also annihilated by the jump operators. The number of stationary states does not depend on $\delta$.}.
Here we focus on the two uniform fully polarized states $\ket{\Uparrow}=\bigotimes_{j=1}^N\ket{\uparrow}$ and $\ket{\Downarrow}=\bigotimes_{j=1}^N\ket{\downarrow}$, as they are stationary for every $N$, regardless of the boundary conditions,
and they survive the injection of fluctuations (see discussion below).

As we show in the End Matter, an arbitrary small amount of classical fluctuations, $T>0$, leads to the formation of a unique mixed steady-state as soon as correlations beyond mean-field are included, regardless of $\delta$.
On the contrary, bistability is robust with respect to quantum fluctuations, and survives up to a $\delta$-dependent critical value $\Omega_c$. 
The fEM, then, showcases the different effect of classical and quantum fluctuations, providing an example of collective dynamics fragile to the first and robust to the second, as opposed to similar models in two dimensions~\cite{kazemi2021genuine,brighi2026dissipative}. 

\begin{figure}[t]
    \centering
    \includegraphics[width=1.0\linewidth]{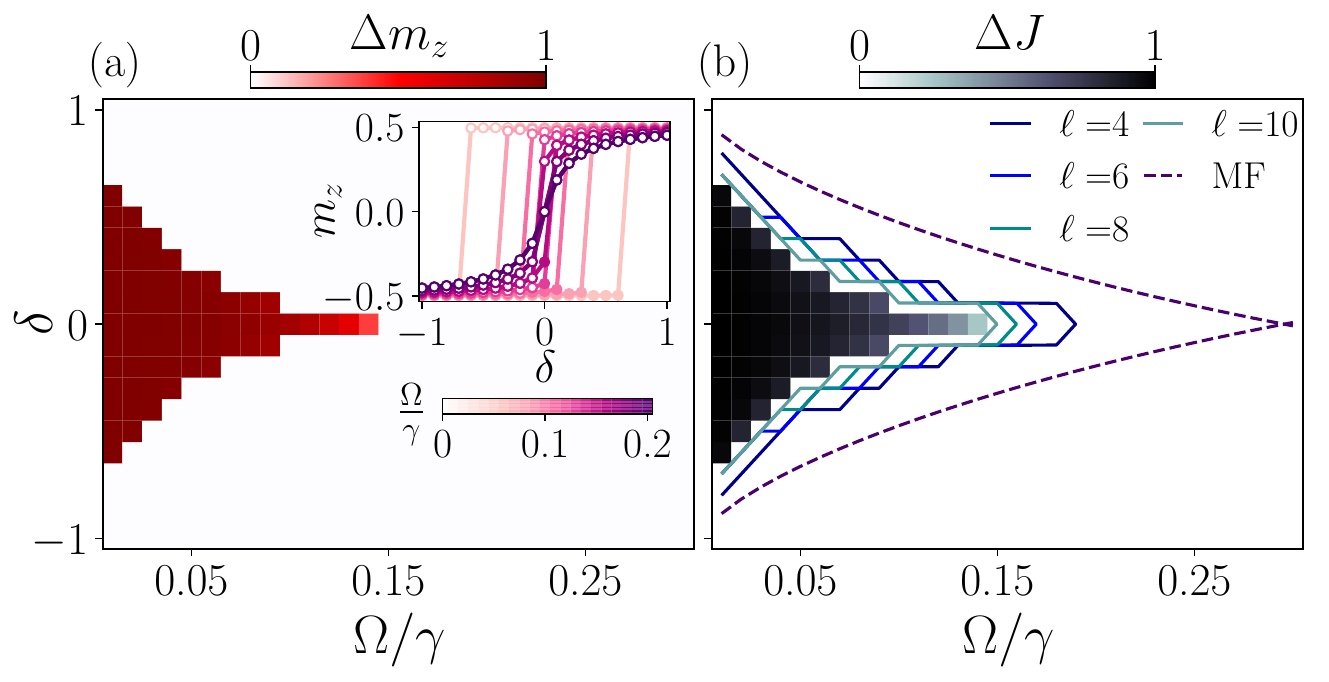}
    \caption{\label{Fig:cmf phase diag}
    (a): Phase diagram for cluster size $\ell=10$ obtained through the hysteresis cycle of the global magnetization (inset).
    The forward and backward sweep have opposite magnetization in the bistable phase, resulting in finite $\Delta m_z$, sharply vanishing at the phase boundary $\Omega_c(\delta)$.
    (b): Similar features are observed for the emission rates, indicating that bistability affects the dynamical behavior of trajectories.
    Solid lines represent the phase boundaries $\Omega_c(\delta)$ obtained for different cluster sizes $\ell$, while the dashed line is the mean-field result, immediately shrinking as correlations are included.
    }
\end{figure}

We now explore quantitatively the structure of the steady-state phase diagram in the $(\Omega/\gamma,\delta,T=0)$ parameter space, looking for signatures of bistability ($T>0$ results can be found in the End Matter).
First, we compute the mean-field equations of motion, which provide insightful information on the behavior of the system. 
To this aim we consider a factorized ansatz for the system density matrix $\rho=\bigotimes_j\rho_j$ and assume translational invariance ($\rho_j=\rho_{\rm MF}, \forall j$).
This results in a set of coupled non-linear differential equations for the local magnetization $m_\alpha=\tr[\Sa{j}\rho_{\rm MF}]$ with $\alpha=x,y,z$
\begin{align}
    \dot{m}_x &= -\gamma \left(\frac{3}{8}+\frac{\delta}{2} m_z-\frac{1}{2}m^2_z\right)m_x \label{Eq:mf_mx}\\
    \dot{m}_y &= -\Omega m_z - \gamma\left( \frac{3}{8} + \frac{\delta}{2}m_z - \frac{1}{2}m_z^2 \right)m_y \label{Eq:mf_my}\\
    \dot{m}_z &= \Omega m_y + \gamma\frac{\delta}{8} +\gamma\left(\frac{1}{4} - \frac{\delta}{2}m_z - m_z^2\right)m_z \label{Eq:mf_mz}.
\end{align}
Crucially, the kinetic constraints in the jump operators result in cubic terms in the mean-field equations and allow multiple stable solutions for the steady-state. Solving for $\dot{m}_\alpha = 0$ at fixed $\gamma$, we find two stable solution when $|\delta|<|\delta_c(\Omega)|\approx\left[1-\frac{5}{2}\left(\Omega/\gamma\right)^{2/3}+\frac{5}{4}\left(\Omega/\gamma\right)^2\right]$, corresponding to the region within the dashed line in Figure~\ref{Fig:cmf phase diag}(b).
On the $\delta = 0$ line one can get an exact evaluation of the critical transverse field $\Omega_c=\sqrt{3/32}\gamma$. 
We find that the remnant magnetization vanishes as $m_z\propto \pm|\Omega-\Omega_c|^\beta$ with a critical exponent $\beta = \frac{1}{2}$, suggesting that the mean-field phase transition might belong to the Ising universality class.

We can refine this analysis using cluster mean-field (CMF)~\cite{jin2016cluster,jin2018phase}, consisting of an improved Gutzwiller ansatz, where the global density matrix gets factorized into identical clusters $\mathcal{C}$ of size $\ell$, $\rho = \otimes_\mathcal{C}\rho_\mathcal{C}$. This allows to include correlations exactly within the single cluster, and treat only longer range correlations at the mean-field level. Assuming translational invariance of the clusters directly probes the thermodynamic limit, where the action of sites outside of the cluster is accounted for by their mean-field contribution.
Using the CMF ansatz, we probe bistability performing a forward and backward sweep of the parameter $\delta$.
In the forward sweep, we start from $\delta = -1$, reach the steady-state, and use it as initial state for the next value $\delta_\text{next} = \delta_\text{prev} + d\delta$, with $d\delta = 0.1$.
We proceed until we reach $\delta = 1$ and then reverse the process and perform the backward sweep, $d\delta \to -d\delta$.

Bistability manifests as an hysteresis cycle in the global magnetization $m_z = \frac{1}{\ell}\sum_j \langle \Sz{j}\rangle_{\rm ss}$ (where $\langle\bullet\rangle_{\rm ss}\equiv\lim_{t\to\infty}\tr[\bullet\rho_{\mathcal{C}}(t)]$), as shown in the inset of Fig.~\ref{Fig:cmf phase diag}(a).
In the forward sweep, $m_z$ remains negative even past $\delta = 0$, while in the backward sweep it remains positive at $\delta < 0$.
Therefore, we define the difference in the forward and backward magnetization as a suitable order parameter for bistability
\be
\label{Eq:Delta mz}
\Delta m_z = |m_z^{(f)} - m_z^{(b)}|.
\ee
As we show in Figure~\ref{Fig:cmf phase diag}(a), in presence of quantum fluctuations $\Delta m_z>0$ within a finite region in parameter space, highlighting the emergence of a robust bistable phase.
By construction, the critical exponent $\beta=1/2$ retains its mean-field value within the CMF framework.

Bistability is also manifest in the statistics of emission in quantum trajectories. The relevant order parameter is given by the (normalised) difference in average number of emissions between the forward and backward sweeps:
\begin{equation}
    \label{Eq:Delta J}
    \Delta J = \frac{|\langle\hat L_{-}\rangle^{(f)}_{\rm ss}-\langle \hat L_{-}\rangle^{(b)}_{\rm ss}|}{\langle \hat L_{+}\rangle^{(f)}_{\rm ss}+\langle \hat L_{+}\rangle^{(b)}_{\rm ss}},
\end{equation}
where $\hat L_{\pm}= \sum_j \left( \left(\hat{L}^{\uparrow\uparrow\uparrow}_j\right)^\dagger \hat{L}^{\uparrow\uparrow\uparrow}_j \pm \left(\hat{L}^{\downarrow\downarrow\downarrow}_j\right) ^\dagger\hat{L}^{\downarrow\downarrow\downarrow}_j \right)$. 
Since the operators $(\hat{L}^{\uparrow\uparrow\uparrow/\downarrow\downarrow\downarrow}_j)^\dagger \hat{L}^{\uparrow\uparrow\uparrow/\downarrow\downarrow\downarrow}_j = \mathbb{P}^{\downarrow/\uparrow}_{j}\mathbb{P}^{\uparrow/\downarrow}_{j+1}\mathbb{P}^{\uparrow/\downarrow}_{j+2}$ are products over three sites, $\Delta J$ is a higher-order correlation function. As shown in Fig.~\ref{Fig:cmf phase diag}(b), the phase diagram in terms of $\Delta J$ mirrors that of the magnetization. 
Finally, comparing the critical line for different cluster sizes, we notice that the bistable phase remains finite and that its boundary shrinks as $\ell$ increases. By plotting the critical field $\Omega_{\rm c}(\delta)$ against $1/\ell$ and extrapolating to $1/\ell\to0$, we find that $\Omega_{\rm c}(\delta)>0$ in a finite region around $\delta=0$, with $\Omega_{\rm c}\approx0.15\,\gamma$ at $\delta=0$.

\textbf{\textit{Relaxation dynamics--}}
The CMF results suggest the robustness of the bistable phase.
To further characterize the two phases through their dynamical properties, we perform exact numerical simulations using a Monte Carlo matrix product state approach \cite{maki2023monte} which combines standard time-evolving block decimation~\cite{vidal2003efficient} and stochastic trajectory dynamics \footnote{ Due to the $3$-site nature of the dissipators we need to slightly modify the usual TEBD.}. 
In this way we can probe the trajectory-resolved dynamics of systems of hundreds of sites and explore the dynamical behavior of the bistable and normal phases. In particular we consider quantum jumps unravellings of the master equation \cite{Molmer1993,plenio1998the-quantum-jump,daley2014quantum}. 

\begin{figure}[t]
    \centering
    \includegraphics[width=0.99\linewidth]{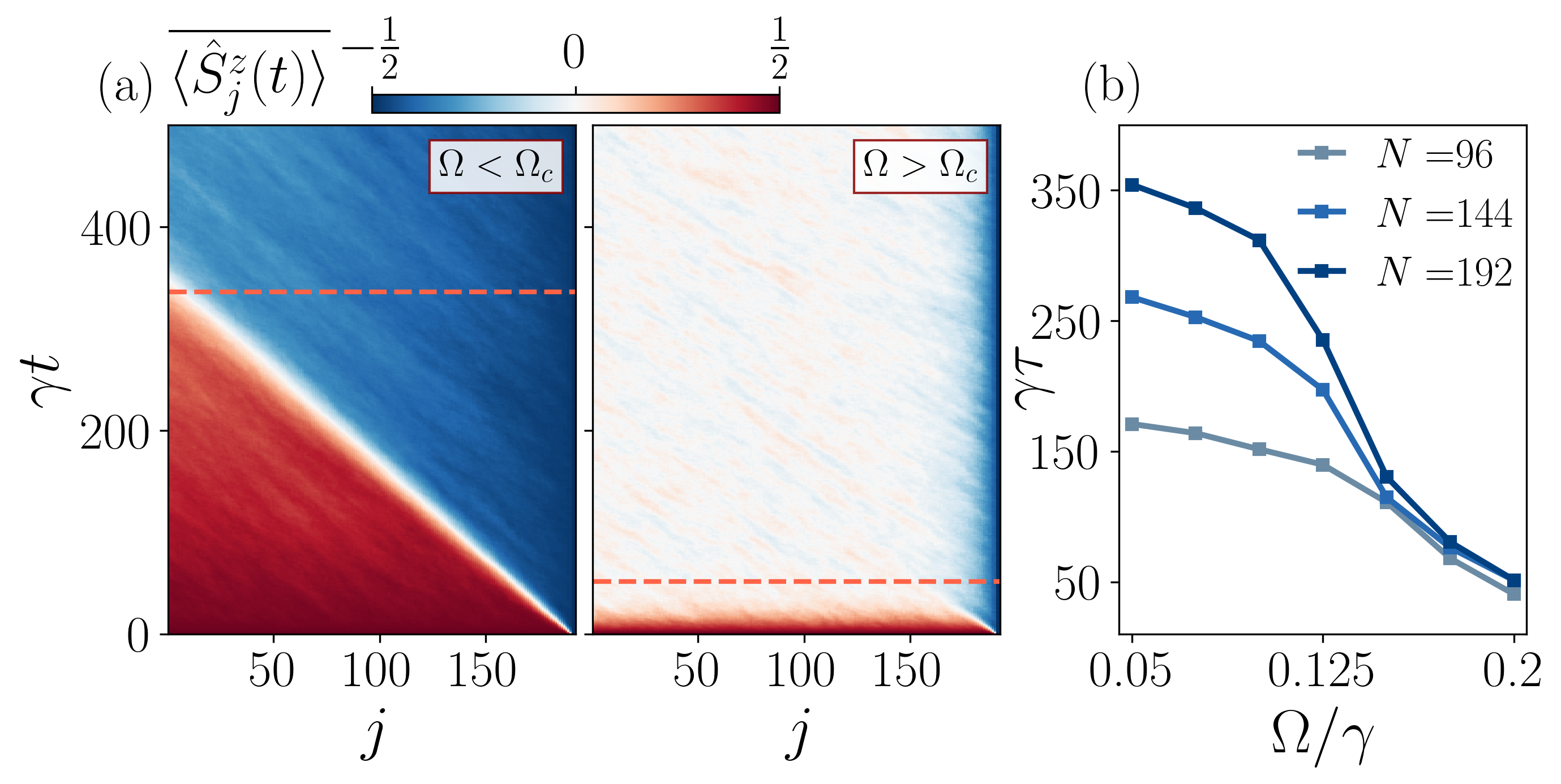}
    \caption{\label{Fig:average Sz + tau}
    (a): Average local magnetization dynamics.
    In the bistable phase (left, $\Omega = 0.075\gamma$), the boundary between up and down spins is well defined and propagates ballistically to the left edge of the chain.
    On the other hand, in the normal phase (right, $\Omega = 0.2\gamma$), local magnetization is quickly scrambled across both space and time.
    (b): Relaxation time of $\ket{\psi_{\rm DW}}$ across the phase transition.
    Bistability results in $\tau\propto N$; at large $\Omega\geq 0.15\gamma$, instead, the system quickly reaches the steady-state, irrespective of system size.
    Data are obtained averaging over $\mathcal{N} = 500$ trajectories.
    }
\end{figure}

We consider the evolution under open boundary conditions of domain-wall initial states of the form
\be
\label{Eq:psi0}
\ket{\psi_{\rm DW}} = \ket{\overbrace{\uparrow\uparrow\dots\uparrow}^{\ell_\uparrow}\underbrace{\downarrow\downarrow\dots\downarrow}_{N-\ell_\uparrow}}.
\ee
This choice is useful to probe the phase transition in large systems, as dynamics are clearly different in the two phases~\cite{brighi2026dissipative}.

In Fig.~\ref{Fig:average Sz + tau}(a) we show the local magnetization dynamics for an initial state with a large \textit{island} of positive magnetization $\ell_\uparrow = N-2$ (and thus $N-\ell_\uparrow=2$) for $N = 192$ at $\delta = 0$.
In the bistable phase (left panel), we observe that the initially large positive magnetization region is progressively eroded and the boundary between up and down spins remains well defined and propagates ballistically to the left as an effect of chiral dissipation. As a result the system equilibrates towards a uniform state with all spin pointing down on a timescale $\tau$ (dashed line) that scales linearly with $\ell_\uparrow$. By choosing $\ell_\uparrow\propto N$ it is clear that we set initial states that relax on a timescale that diverges with the system size as $\tau\propto N$. In the normal phase (right panel) the domain mixes very quickly with the neighboring spins, and quantum fluctuations wash out the characteristic boundary dynamics typical of the bistable phase, resulting in rapid melting towards a disordered state with vanishing magnetization. 

In Fig.~\ref{Fig:average Sz + tau}(b) we show the relaxation time $\tau$, measured as the time when the island disappears, as a function of $\Omega$ for different system sizes and $\ell_\uparrow = N-2$.
There is a clear difference in the behavior of $\tau$ between small and large $\Omega$: in the bistable phase $\tau \propto N$, while above the critical field it becomes independent of system size.
The estimated $\Omega_c\approx 0.15 \gamma$ confirms the CMF analysis.

\begin{figure}[b]
    \centering
    \includegraphics[width=0.99\linewidth]{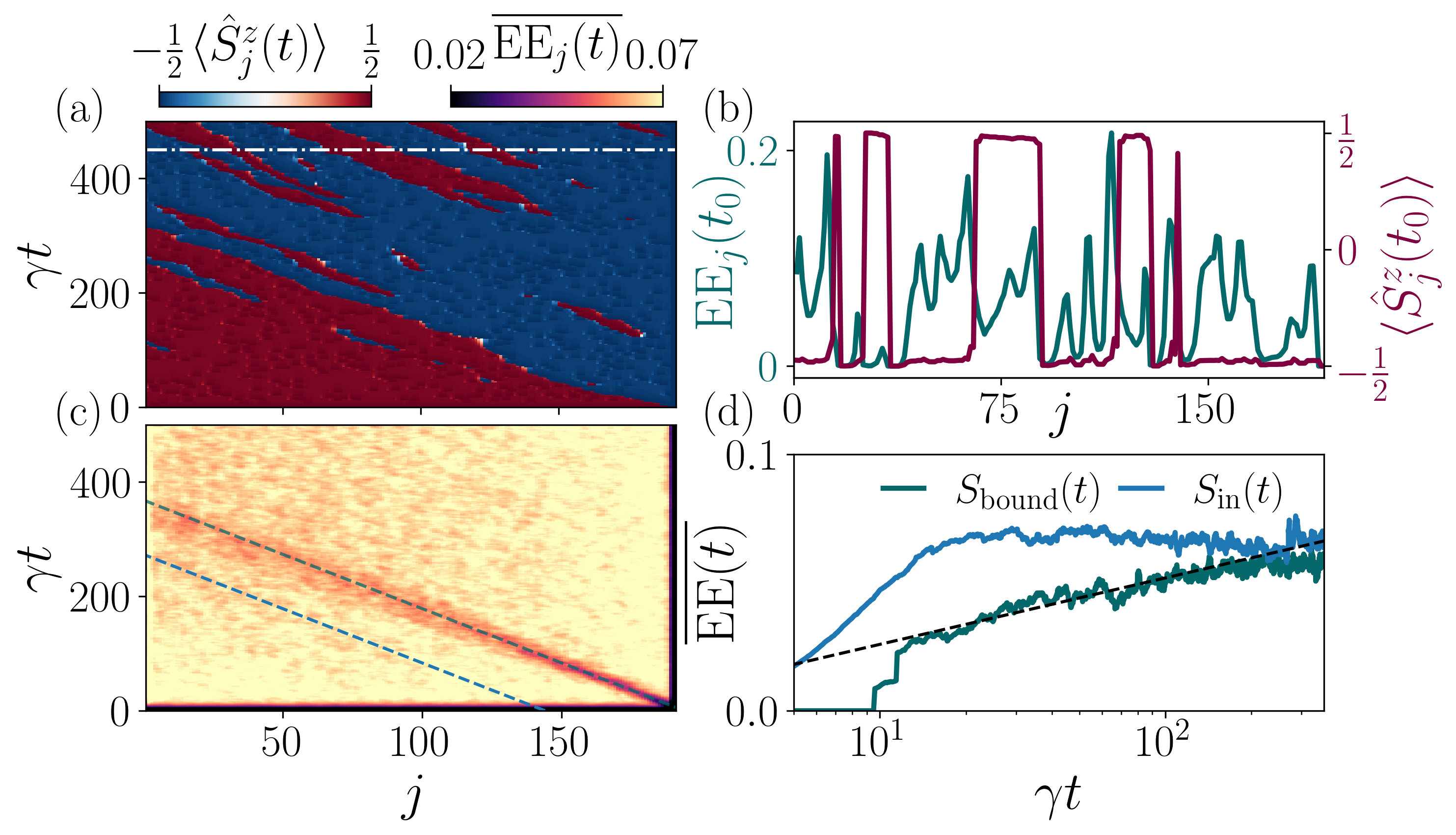}
    \caption{\label{Fig:S + dyn het}
    (a): In the bistable phase, large regions in spacetime where the system is approximately in one of the $2$ steady-states coexist; this is a typical feature of dynamical heterogeneity.
    (b): Magnetization and entanglement entropy profiles at fixed time $ t_0 = 450/\gamma$ [dotted white line in panel (a)].
    Dynamical heterogeneity manifests in the sudden drop of entanglement entropy at the boundary between different magnetization regions.
    (c): Average entanglement dynamics in the bistable phase ($\Omega = 0.075\gamma$) show a clear separation between the two regions inside and outside the large island, divided by a boundary with low entanglement entropy.
    (d): Entanglement entropy inside the island quickly saturates to its steady-state value.
    Along the boundary, instead, entanglement growth is much slower (logarithmic in time).
    Results are obtained averaging over $\mathcal{N} = 500$ trajectories.
    }
\end{figure}

\textbf{\textit{Dynamical heterogeneity in quantum trajectories--}}
The increase of relaxation time with system size indicates that the system evolves in a correlated manner. This becomes evident at the level of single quantum trajectories. We show in Fig.~\ref{Fig:S + dyn het}(a) that individual trajectories display large space-time regions of the opposite magnetization when the system is in the bistable region. These are generated by the quantum fluctuations induced by $\hat{H}$ and would be absent at $\Omega = 0$. This dynamical heterogeneity~\cite{garrahan2002geometrical,causer2025dynamical} is a typical feature of glassy systems, and is strictly related to their long relaxation times and to the presence of constraints. 

Heterogeneity in the dynamics is also made evident in genuine quantum features, such as the growth of entanglement.
For each trajectory $\ket{\psi(t)}$, we evaluate the entanglement entropy $\text{EE}_j$ for a bipartition of the chain across the bond $j$, $\text{EE}_j\!=\!-\tr\!\left[\rho_{A_j}\ln\rho_{A_j}\right],$
where $A_j\!=\!\{1,\ldots,j\}$ and $\rho_{A_j}\!=\!\tr_{\bar A_j}\ket{\psi(t)}\!\bra{\psi(t)}$ with $\bar A_j\!=\!\{j+1,\ldots,N\} $ \footnote{The entanglement entropy defined in the main text is a well defined entanglement measure for the open-system dynamics since for each trajectory the state remains pure throughout the evolution.}. 
In Fig.~\ref{Fig:S + dyn het}(b) we compare the magnetization (red) and entanglement profile (green) at fixed time $t_0 = 450/\gamma$.
The sharp changes in magnetization, separating regions of opposite $\langle \Sz{j}\rangle$, are always correlated with a drop in entanglement entropy, which otherwise grows within the spontaneously formed islands. 
This suggest that there is almost no entanglement between the different regions and thus at fixed time the wavefunction is well approximated by the tensor product of multiple domains with opposite magnetization.

Whenever regions of opposite magnetization are common between many trajectories, this decoupling will be visible in the average entanglement.
This is exactly the case for the initial state we studied above, as we show in Fig.~\ref{Fig:S + dyn het}(c), where a clear entanglement drop separates the two magnetization regions.
We finally investigate the entanglement dynamics in the boundary moving frame.
In particular, we choose $2$ distinct paths in spacetime.
The first one corresponds to a bipartite cut $j(t)$ always inside the island [blue dashed line in panel (c)], while the second follows the boundary between the two magnetization regions [green dashed line in panel (c)].
As we show in panel~(d), entanglement growth is heterogeneous~\cite{zadnik2023slow}.
Inside the island, entanglement quickly saturates to its steady-state value, while along the boundary it grows logarithmically slow and saturates on much longer timescales.

\textbf{\textit{Conclusions--}}
We have shown that the one-dimensional dissipative far-East model hosts a non-equilibrium phase transition, forbidden in equilibrium for short-range interacting systems.
Interestingly, the resulting collective dynamics and bistability are robust to quantum fluctuations, while being fragile to classical noise.
Combining cluster mean-field theory with stochastic tensor-network simulations, we characterized the resulting non-equilibrium phase diagram and identified a quantum fluctuations-driven phase transition from a bistable to a monostable regime. 
At the dynamical level, we showed that the bistable phase is accompanied by slow relaxation, dynamical heterogeneity, and a characteristic spatial organization of entanglement, highlighting how coherent fluctuations fundamentally reshape the collective dynamics generated by kinetic constraints.

Our results open several directions for future research. 
A natural extension is the investigation of quantum jump statistics \cite{salatino2026} and trajectory large deviations \cite{garrahan2010thermodynamics}, which may provide a complementary characterization of the emergent dynamical phases. 
More broadly, it will be important to understand which classes of kinetically constrained dissipative processes can sustain analogous fluctuation-stabilized collective behavior.
Finally, it will be interesting to investigate whether quantum complexity measures, such as non-Gaussianity \cite{Nongauss_diss} and non-stabilizerness \cite{Liu2022}, can provide complementary signatures of the dynamical phases and phase transitions uncovered here.

\begin{acknowledgments}
\textbf{\textit{Acknowledgments--}}
    PB acknowledges support by the Austrian Science Fund (FWF) [Grant Agreement No.~10.55776/ESP9057324].
    JPG acknowledges support from EPSRC grant no.\ EP/V031201/1 and The Leverhulme Trust Grant no.\ RPG-2024-112.
    AB and GF acknowledge financial support from Provincia Autonoma di Trento. 
    IL acknowledges support through the ERC grant OPEN-2QS (Grant No. 101164443), from the Deutsche Forschungsgemeinschaft (DFG, German Research Foundation) through the Research Unit FOR 5413/1, Grant No. 465199066, and through the Research Unit FOR 5522/1, Grant No. 499180199.
    The tensor-network numerical simulations were performed using the ITensor library~\cite{ITensor} on the Austrian Scientific Computing (ASC) infrastructure.
\end{acknowledgments}

\bibliography{bibliography-04092026.bib}

\appendix
\section{End Matter}

\textit{\textbf{Finite $T$ phase diagram--}}
To investigate how classical fluctuations impact the cooperative behavior of the far-East chain, we introduce unbiased classical noise by setting $\gamma_{\downarrow\uparrow\uparrow} = \gamma_{\uparrow\downarrow\downarrow}  = \bar{\gamma}/2$ which defines the dimensionless classical fluctuations strength $T = \bar{\gamma}/\gamma$. By introducing classical fluctuations, the kinetic constraints are locally violated because the jump operators $\hat{L}^{\downarrow\uparrow\uparrow}_j = \sqrt{\gamma_{\downarrow\uparrow\uparrow}}\SM{j}\Pup{j+1}\Pup{j+2}$ and $\hat{L}^{\uparrow\downarrow\downarrow}_j = \sqrt{\gamma_{\uparrow\downarrow\downarrow}}\SP{j}\Pdn{j+1}\Pdn{j+2}$ permit spin flips in the direction opposite to the neighboring majority. 
En passant, one can notice that these jump operators behave in the exact opposite way than the majority jump operators $\hat{L}^{\downarrow\uparrow\uparrow/\uparrow\downarrow\downarrow}_j\propto\left(\hat{L}^{\uparrow\uparrow\uparrow/\downarrow\downarrow\downarrow}_j\right)^\dagger$, with the proportionality constant being $\sqrt{T/(1\pm\delta)}$.

The introduction of classical fluctuations modifies the MF equations from the form of Eqs.(\ref{Eq:mf_mx})-(\ref{Eq:mf_mz}) to:
\begin{widetext}
\begin{align}
    \dot{m}_x &= -\gamma \left[\frac{3}{8}+\frac{\delta}{2} m_z-\frac{1}{2}m^2_z+\frac{3}{4}T\left(\frac{1}{2}+2m_z^2\right)\right]m_x\label{Eq:mf_mx_T}\\
    \dot{m}_y &= -\Omega m_z - \gamma\left[ \frac{3}{8} + \frac{\delta}{2}m_z - \frac{1}{2}m_z^2 +\frac{3}{4}T\left(\frac{1}{2}+2m_z^2\right)\right]m_y \label{Eq:mf_my_T}\\
    \dot{m}_z &= \Omega m_y + \gamma\frac{\delta}{8} +\gamma\left[\frac{1}{4} - \frac{\delta}{2}m_z - m_z^2- T\left(\frac{3}{4}+m_z^2\right)\right]m_z \label{Eq:mf_mz_T}
\end{align} 
\end{widetext}
By solving Eqs.(\ref{Eq:mf_mx_T})-(\ref{Eq:mf_mz_T}) we find that a bistable region is present also for $T>0$. Nevertheless, classical noise contributes to the effective damping with a strength that grows linearly in $T$, while the coherent field appears as a quadratic term with a denominator proportional to the overall dissipation strength $\gamma$, and increasing with $T$. Consequently, larger values of $T$ favor dissipation over coherent dynamics, reducing the parameter region where bistability is expected to be observed. In particular, in the absence of quantum fluctuations $\Omega=0$ equation (\ref{Eq:mf_mz_T}) predicts bistability at $\delta=0$ up to the critical temperature $T_{\rm c}=1/3$.

\begin{figure}[]
    \centering
    \includegraphics[width=1.0\linewidth]{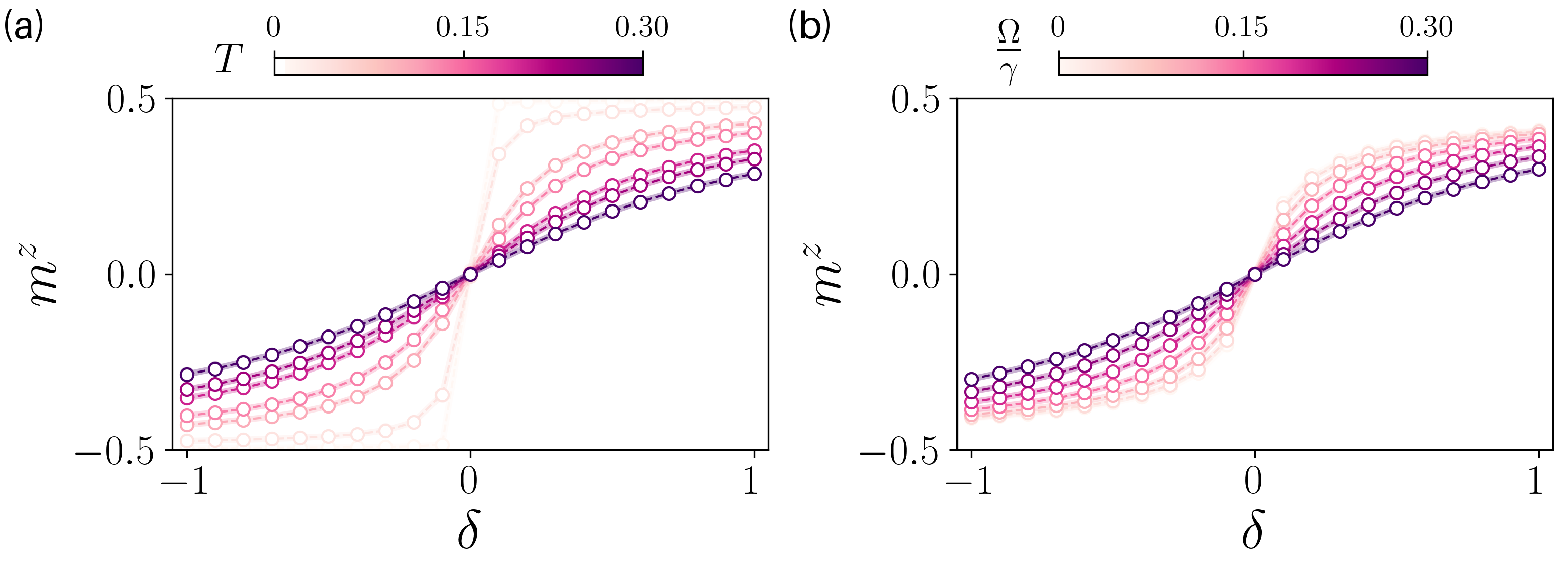}
    \caption{\label{Fig:hysteresis_T} Absence of hysteresis in sweeps of the global magnetization \textit{vs.} $\delta$ (a) at $\Omega=0$ with classical MC simulations and $T\in[0.01,0.30]$; (b) for cluster size $\ell = 10$ at $T=0.2$ and $\Omega\in[0.01,0.30]$. Contrary to the inset of Figure \ref{Fig:cmf phase diag}, the inclusion of classical fluctuations results in the forward and backward sweeps having the same magnetization, implying the absence of bistability.}
\end{figure}

We now go beyond the MF approximation.
In the absence of quantum fluctuations ($\Omega=0$), the Lindblad master equation (\ref{eq:LindME}) strictly reduces to a classical Markov jump process for the spin configurations in the $z$-basis. We exploit this by tracking the exact stochastic evolution via asynchronous Monte Carlo sweeps, meaning we randomly select and update individual spins one at a time according to the local transition rates.
This allows us to compute the phase diagram in the $\delta-T$ plane for very large system sizes $N\sim 16000$ ~\cite{hu2019taichi}, which allows to probe the effect of very weak $T$.
We find that for any probed $T>0$ the bistable phase disappears in favor of the monostable one.
This is illustrated in Fig. \ref{Fig:hysteresis_T}(a), where we demonstrate the absence of hysteresis cycles.

%To better understand the full picture, one can also reintroduce quantum fluctuations. Using the MF equations as a starting point, we expect the introduction of any $\Omega>0$ on top of classical noise to have a further detrimental effect.

When quantum fluctuations are also present ($\Omega>0$) we go beyond the MF introducing correlations through a CMF analysis.
We observe that, contrary to the case of quantum fluctuations alone ($T=0$) analyzed in the main text, the bistability region quickly collapses as cluster size increases as we turn on classical fluctuations. To exemplify this, we show in Fig. \ref{Fig:hysteresis_T} (b) the magnetization sweeps that again, do not show an hysteretic behaviour.
Therefore we can conclude that even in this case the presence of classical fluctuations will not allow for bistability and thus suppress the collective dynamics. 

\end{document}